# Paramagnetic centers in graphene nanoribbons prepared from longitudinal unzipping of carbon nanotubes

S. S. Rao[*] and A. Stesmans

Department of Physics, University of Leuven, Celestijnenlaan 200 D, B-3001 Leuven

INPAC-Institute for Nanoscale Physics and Chemistry, Leuven, Belgium

D. V. Kosynkin[1], A. Higginbotham[1] and J. M. Tour[1,2,3]

[1]Department of Chemistry, [2]Department of Mechanical Engineering and Materials Science, [3]Smalley Institute for Nanoscale Science and Technology, Rice University, MS-222, 6100.

* Srinivasarao.singamaneni@fys.kuleuven.be

Abstract

Electron spin resonance (ESR) investigation of graphene nanoribbons (GNRs) prepared through longitudinal unzipping of multiwalled carbon nanotubes (MWCNTs) indicates the presence of C-related dangling bond centers, exhibiting paramagnetic features. ESR signal broadening from pristine or oxidized graphene nanoribbons (OGNRs) is explained in terms of unresolved hyperfine structure, and in the case of reduced GNRs (RGNRs), the broadening of ESR signal can be due to enhancement in conductivity upon reduction. The spin dynamics observed from ESR linewidth-temperature data reflect a variable range hopping (VRH) mechanism through localized states, consistent with resistance-temperature data.

The recent interest in the physics and applications of graphene nanoribbons (GNRs), with tunable band gap over graphene and higher carrier mobilities over carbon nanotubes (CNTs) and Si, for future generations of nanoelectronic devices has evoked impressive research. Field-effect transistors (FETs) based on GNRs have already been demonstrated[1]. Several proposals have been made to apply GNRs in high magneto-resistive[2] and high frequency devices[3].

A noteworthy feature of GNRs is the appearance of 'edge-spin' polarization at the edges of the zigzag-edged GNRs (ZGNRs). Recently[4], it has been shown that such edge-spin polarization can be electrically controlled to induce a half-metallic band structure. Therefore, heading towards the realization of GNRs spintronic devices, it is important to understand the spin dependent properties of ZGNRs. Theoretical work[5] has shown that ZGNRs are magnetic and can carry a spin current in the presence of a sufficiently large electric field.

Theoretical studies[6] showed the presence of σ- and π- defects at GNR edges, a property that was proposed for building spin filters. Using first-principle calculations[7], the effect of magnetic point defects (vacancies and adatoms) was investigated in ZGNRs. While pristine ribbons display anti-parallel spin states at their edges leading to anti-ferromagnetism, the defects are found to perturb this coupling. The introduction of a vacancy drastically reduces the energy difference between parallel and anti-parallel spin orientations, though the latter is still favored. Moreover, the local magnetic moment of the defect is screened by the edges so that the total magnetic moment is quite small. In contrast, when an adatom is introduced, the parallel spin orientation is preferred and the local magnetic moment of the defect adds up to the contributions of the edges.

Furthermore, a spin-polarized transmission is observed at the Fermi energy, suggesting the possible use of such GNRs in spin-valve devices. Recent theoretical calculations[8] predicted that the total spin moment of a GNR should be zero as the local magnetic moments at the two edges are coupled anti-ferromagnetically. In the same study, it was also concluded that hydrogenated GNRs can have a finite total magnetic moment leading to a ferromagnetic ground state. Hence, it appears that investigating the role of defects is highly important in realizing GNRs in digital electronics and spintronics. Several research groups have demonstrated experimentally ferromagnetic-like features in graphite[9], proton irradiated graphite[10], carbon films[11], and $C_{60}$-based polymers[12]. Experimental work probing the magnetic features in graphene, a building block of graphite has been scarce[13]. The observed magnetism was shown to originate mainly from isolated vacancies, vacancy-hydrogen complexes, grain-boundaries, and planar and topological line defects, as these defects possess localized spin moments and thus may lead to magnetism.

Despite a plethora of theoretical studies and numerical calculations which describe the exotic properties of graphene and GNRs, in particular in regard to the presence of defects and their magnetic properties, it appears that little experimental work has been carried out on the nature of defects in graphene[14,15]. Progress[16] in producing substantial quantities of high quality GNRs using a chemical route has enabled and motivated us to perform first electron spin resonance (ESR) measurements to gain detailed information and to test theoretical predictions. Even though electronic transport measurements performed on monolayer GNR devices showed an ambipolar electric field effect typical for graphene, the conductivity of monolayer GNRs (~35 S/cm) and mobility of charge carriers (0.5-3 cm$^2$/Vs) are found to be less than those of pristine graphene. The reason for inferior values was explained[15] as a result of harsh oxidative treatment during the unzipping of MWCNTs, which produces defects in GNRs thereby leading to a decrease in conductivity and mobility. Accordingly, the main goal of the present work is two-fold: a) to probe by ESR the charge trapping centers causing the lower conductivity (lower mobility) in comparison with graphene and graphite exfoliated ribbons; b) to investigate the magnetic nature and spin dynamics of these defects On the basis of ESR probing, a suitable local probe for atomic identification of defects and their local surroundings, we suggest that the observed magnetism originates from C-related dangling bond defects which form localized states.

GNRs were prepared by chemical longitudinal unzipping of multi-walled carbon nanotubes (MWCNTs) as reported by Kosynkin *et al*[16]. Briefly, this method involves the treatment of MWCNTs, consisting of 15–20 concentric cylinders, of 40–80 nm diameter with concentrated $H_2SO_4$ followed by $KMnO_4$ (an oxidizing agent) at room temperature. This process chemically unzips the nanotubes, forming nanoribbons up to 4 μm long, with widths of 100–500 nm and thicknesses of 1–30 graphene layers: these samples are labeled as oxidized GNRs (OGNRs). These were characterized by several techniques such as X-ray diffraction (XRD), scanning electron microscopy (SEM), transmission electron microscopy (TEM), atomic force microscopy (AFM), thermogravimetric analysis (TGA), atenuated total reflection infra red (ATR-IR) spectroscopy, and X-ray photoelectron spectroscopy (XPS). Electronic properties of the thus obtained monolayer GNRs have also been measured[17], exhibiting FET characteristics, and the temperature dependent electronic transport obeyed a variable range hopping (VRH) mechanism. It has been found that OGNRs (as prepared) are poor conductors due to the disruption of π conjugation. The conductivity could be improved significantly after reduction with $N_2H_4$ resulting in partial restoring of the disrupted conjugated π network – samples labeled as reduced GNRs (RGNRs). More details on the formation mechanism of GNRs, characterization and physical properties can be found in Refs. 16,17. In the current work, to investigate the effect of $H_2$ on OGNRs and in an attempt to passivate C-related defects to enhance conductivity and mobility, OGNRs were treated in $H_2$ at 350 °C (1.1 atm) for 37 min, this sample is now being labeled as hydrogenated GNRs (HGNRs).

Conventional first derivative CW ESR experiments were carried out on OGNRs, RGNRs and HGNRs in the T range of 4.2 – 120 K using a home built K-band (≈ 20.6 GHz)[18] and Q-band ( ≈ 34 GHz) Bruker EMX spectrometer. All were operated under conditions of adiabatic slow passage. Conventional low power first-derivative-absorption $dP_{\mu r}/dB$ ($P_{\mu r}$ being the reflected microwave power) spectra were detected through applying sinusoidal modulation (~ 100 kHz, amplitude $B_m$ ~ 0.52G) of the externally applied magnetic field $\vec{B}$, with incident microwave power $P_\mu$ as well as $B_m$ cautiously reduced to avoid signal distortion. The defect spin density was quantified by double numerical integration of the K-band derivative absorption spectra by making use of a co-mounted calibrated Si:P intensity marker, also serving as g marker: g(4.2 K = 1.99869). Obtained relative and absolute accuracies on defect densities are estimated at ~ 5% and 12%, respectively.

Figure 1 shows representative low-power K-band ESR spectra measured on samples OGNRs (a), RGNRs (b) and HGNRs (c) at 4.2 K. As shown, only one ESR signal, of symmetric shape is observed at zero–crossing g value ($g_c$) = 2.0029, 2.0032, 2.0031, respectively. This value falls within the reported [19] carbon ESR signal range (g = 2.0022 – 2.0035), indicating the signal and may be ascribed to C related dangling bonds of spin S = ½. For all three samples, the 4.2 K signal can be fitted by a Lorentzian lineshape (closely) of peak-to-peak width of $\Delta B_{pp}$ = 2.8, 6.4 and 4.3 G and spin densities of $\approx 2.3 \times 10^{18}$, $1.5 \times 10^{17}$ and $2 \times 10^{17}$ spins/g were inferred for OGNRs, RGNRs and HGNRs respectively. The ESR spectral parameters have been inferred at various T's by means of the computer simulations using the Lorentzian line shapes with optimized fitting parameters. Despite intense signal averaging over broad field ranges under various extreme and optimized spectrometer parameter settings, no other signals could be observed over a broad magnetic field sweep range up to 9000 G. Though intensely searched for, neither any correlated additional signal structure could be traced nor any sign of hyperfine structure possible ensuing from highly abundant $^{1}H$, $^{14}N$, $^{16}S$ and $^{55}Mn$ nuclei.

Figure 2 shows the temperature dependences of $\Delta B_{pp}$, g, and the integrated intensity (proportional to $\chi_{ESR}$) for OGNRs. We find that $\Delta B_{pp}$ decreases slowly with the increase of T, while g is found to be almost constant at $g_{av}$=2.0029±0.0001; ESR spin susceptibility ($\chi_{ESR}$) exhibit a Curie-like paramagnetic behavior. The average g value ≈ 2.0029 is almost constant throughout the T range investigated. The C-based paramagnetic centre may arise from oxygen containing functionalities such as carbonyls and carboxyls residing at the edges and/or surface.

The inferred T dependences of $\Delta B_{pp}$, g and $\chi_{ESR}$ of RGNRs, shown in Fig. 3, are very similar to those of OGNRs (Fig. 2). This may indicate the ESR signal to originate from the carboxylic groups which could not be reduced by $N_2H_4$, as the latter treatment would eliminate most of the carbonyl groups [16].

As indicated above, defect densities remained almost unaltered after treatment of OGNRs in $H_2$. Yet, we noticed a substantial increase in the conductivity. Indeed, no ESR measurements could be performed because of excessive loading of the cavity Q-factor for $T \geq 20$ K (not shown) due to the developed semiconducting properties of GNRs as a result of the $H_2$ treatment. Besides, after this treatment, other ESR properties were left unaffected.

As evident from Figs. 2 and 3 the g value and $\Delta B_{pp}$ increase upon $N_2H_4$ treatment of OGNRs. $\Delta B_{pp}$ increases by more than a factor of 2 (at 4.2 K). This can not have resulted from enhanced dipole – dipole (DD) interaction as the defect density is observed to decrease by an order of magnitude upon treatment in $N_2H_4$. $\Delta B_{pp} \sim 2.8$ G can be explained by invoking unresolved hyperfine broadening, which is temperature independent. One possibility is that spin clustering could also locally increase the spin density and hence the dipolar contribution to $\Delta B_{pp}$.

On reduction of GNRs by $N_2H_4$, the oxygen containing functional groups decreased (though not removed completely), leading to enhancement in conductivity and thereby decreasing the ESR "visible" spin density by an order of magnitude. The estimated DD contribution in RGNRs to $\Delta B_{pp}$ is 0.024 G, negligibly small. According to previous suggestions[20], exchange narrowing cannot be a significant $\Delta B_{pp}$ determining factor given that $\Delta B_{pp}$ increases as the spin concentration decreases. Hence, the enhancement in the conductivity could be the main plausible candidate for $\Delta B_{pp}$ and also for the enhancement in the observed g value. A cross sectional TEM image of monolayer

OGNR is shown in the Fig. 4 of Ref [17] illustrating the non-uniform disorder structure due to the presence of functional groups, which may lend support for the observation of localized states. The high density of "ESR-visible" C-defects may arise from the sizeable disorder present in the GNR. In light of decrease in resistance upon reduction treatment as well as also from earlier suggestions [] in which unpaired electrons interact with conduction electrons (ESR "invisible") to shorten spin relaxation times and therefore increase in $\Delta B_{pp}$, we suggest that enhancement in conductivity upon reduction treatment of GNRs is the dominant $\Delta B_{pp}$ determining factor. The effect of molecular paramagnetic oxygen in broadening the ESR signal can not be ruled out either, which forms the subject of future work.

The information related to spin dynamics might come from the functional dependence of $\Delta B_{pp}$ with T. According to the VRH mechanism[21], in the paramagnetic phase, the T dependence of $\Delta B_{pp}$ can be described by $\Delta B_{pp} = K \exp(\frac{T_0}{T})^{1/n}$ where n-1 is the dimensionality of the system and $T_0$ and K are constants and in analogy with the T dependence of resistance (R)[17]. Figure 4 describes the parallelism between the T (10 – 120 K) dependence of Q-band ESR $\Delta B_{pp}$ (left Y-axis) as observed from on an ensemble of freestanding RGNRs and the T (35 – 120 K) dependence of R (right Y-axis) extracted from a monolayer GNR held by the Si substrate. It appears that this dependence qualitatively follows VRH mechanism which exhibits linear relationship with $\ln(\Delta B_{pp})$ versus $T^{-1/3}$ coordinates for a two-dimensional (2D) system as shown in the inset of Fig. 4, thus further supporting the existence of localized states, created by atomic defects at the graphene edges in GNRs. Hence, from these measurements, it is inferred that the charge transport is dominated by hoping through localized states. The VRH type electron transport mechanism has also been reported[22] in lithographically patterned disordered GNR.

Another interesting observation is the increase of g value from the oxidized state (OGNRs) to reduced state (RGNRs). In the oxidation procedure, $H_2SO_4$ was used to obtain GNRs. During that procedure, there may be entrapement of $HSO_4^-$ acceptors between the graphite layers that may be responsible for the reduction of g value. After elimination of such acceptors upon the reduction, which also leads enhancement in conductivity, could cause an increase in g value. With a first ESR investigation performed on GNRs, the obvious quest concerns the atomic nature of the defect at the origin. The chemically bonded groups such as - OH, -O, -CO, -C=O can form ESR radicals. The g-factor ~ 2.00306 of the benzoin radical[23] (-(CO)O) is close to the observed g factor of RGNRs.

Recently[14] reported temperature-dependent ESR data on mechanically exfoliated graphene indicated conical band dispersion (linear $\chi$ (T)) as expected for graphene, which appears to be absent in the current GNRs due to the presence of non-conducting oxygen containing functional moieties. In the same study very weak-T dependence of $\Delta B_{pp}$ is also reported at high T (> 50 K), along with an increase in linewidth at low T (< 50 K) due to defects. However, the $\Delta B_{pp}$ (9 G) and g value (2.004) measured at 4 K are found to be higher than those observed here. Even though, in the current sample, the strong T dependence of resistance (R) is observed for monolayer GNR, the T dependence of $\Delta B_{pp}$- indicative of spin dynamics is much weaker, partially, could be due to cumulative effect of grouped GNRs.

To compare ESR spectral parameters with those of GNRs, we have also measured ESR on highly oriented pyrolytic graphite (HOPG) and MWCNTs samples. We find a highly anisotropic Dysonian shaped ESR signal and an isotropic Lorentzian shaped itinerant ESR signal as expected from HOPG and MWCNTs, respectively. The electronic properties of GNRs are found to be distinctly different from those of MWCNTs and HOPG.

To summarize, the current ESR results demonstrate that GNRs, produced from longitudinal unzipping of MWCNTs suffer from C-related localized states, exhibiting paramagnetic features. These are believed to be potential charge trapping centers leading to the decrease in conductivity and mobility of GNRs, that is, making them inferior to pristine graphene. The spin dynamics of GNRs appear to follow a VRH mechanism through localized states, supporting the electronic transport data.

The work at Rice University was supported by the Air Force Office of Scientific Research FA9550-09-1-0581 and the Office of Naval Research MURI Graphene program.

**Figure Captions:**

1. First derivative CW K-band ESR spectra measured at 4.2 K on (a) OGNRs (b) RGNRs and (c) HGNRs, subjected to different post-manufacturing treatments: using $B_m$ = 0.42 G and $P_\mu$ = 2.5 nW. The signal at g ≈ 1.99869 stems from a co-mounted Si:P marker sample

2. The temperature dependences of K-band ESR spectral parameters obtained on OGNRs using optimized computer simulations revealing an almost Lorentzian lineshape. The solid line in Fig. 2 (c)guides the eye.

3. The temperature dependences of K-band ESR spectral parameters obtained on RGNRs using optimized computer simulations using almost Lorentzian lineshapes. The solid line in Fig. 3 (c)guides the eye.

4. The T-dependence of Q-band (~ 34 GHz) ESR $\Delta B_{pp}$ of freestanding ensemble of RGNRs (left Y-axis), obtained through computer simulations using Lorentian line shapes. The T (35 – 120 K)-dependent resistance (R) measured on monolayer GNRs on a Si substrate is also shown on right Y-axis [17]. The close parallelism between the $\Delta B_{pp}$ (T) and R (T) dependences, expressed in log scale, hints that the VRH mechanism, describing the R (T) data, also applies for the $\Delta B_{pp}$ (T) results. This is further exemplified in the inset, exposing a linear $\ln(\Delta B_{pp})$ versus $T^{-1/3}$ relationship for 2D system. The solid line guides the eye.

**Figures:**

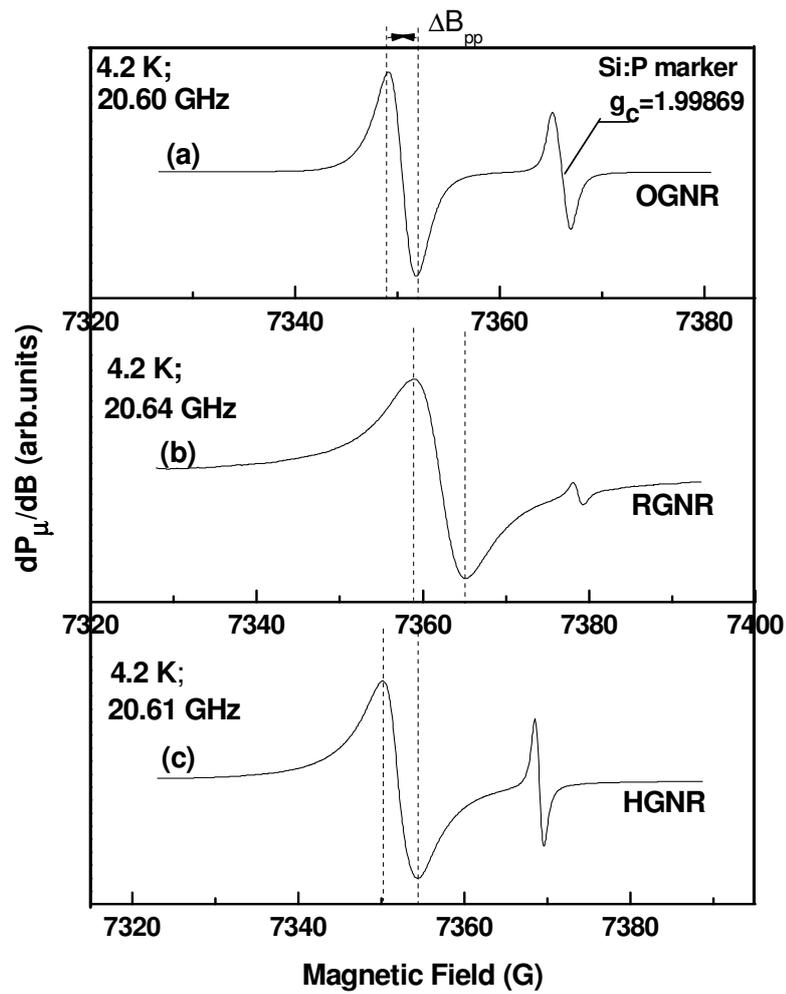

**Fig. 1**

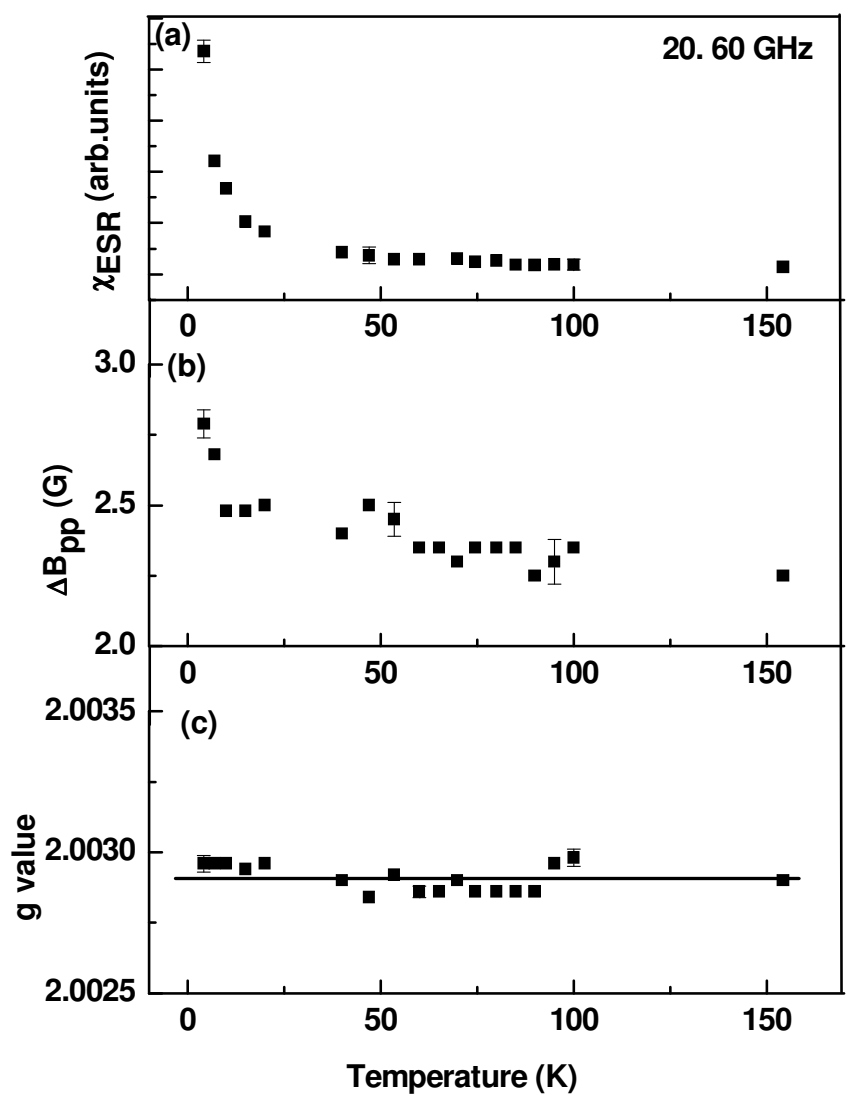

**Fig. 2**

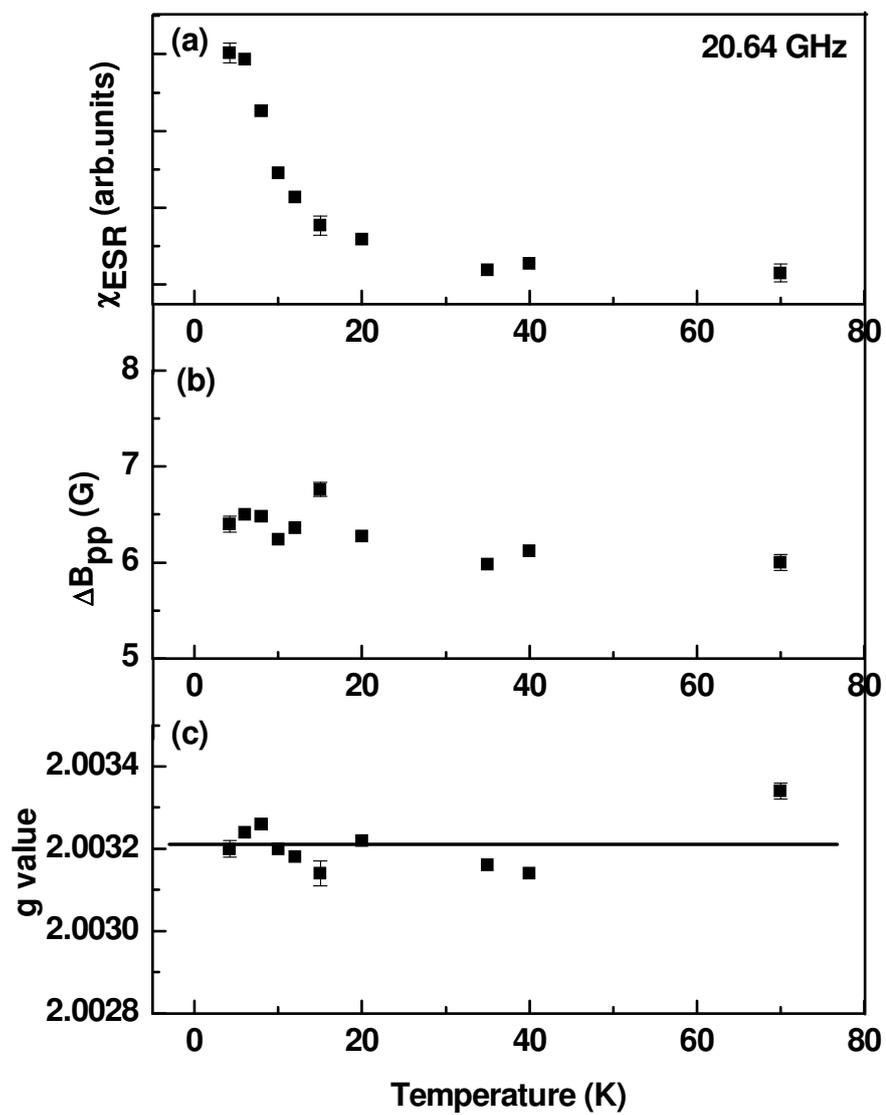

**Fig. 3**

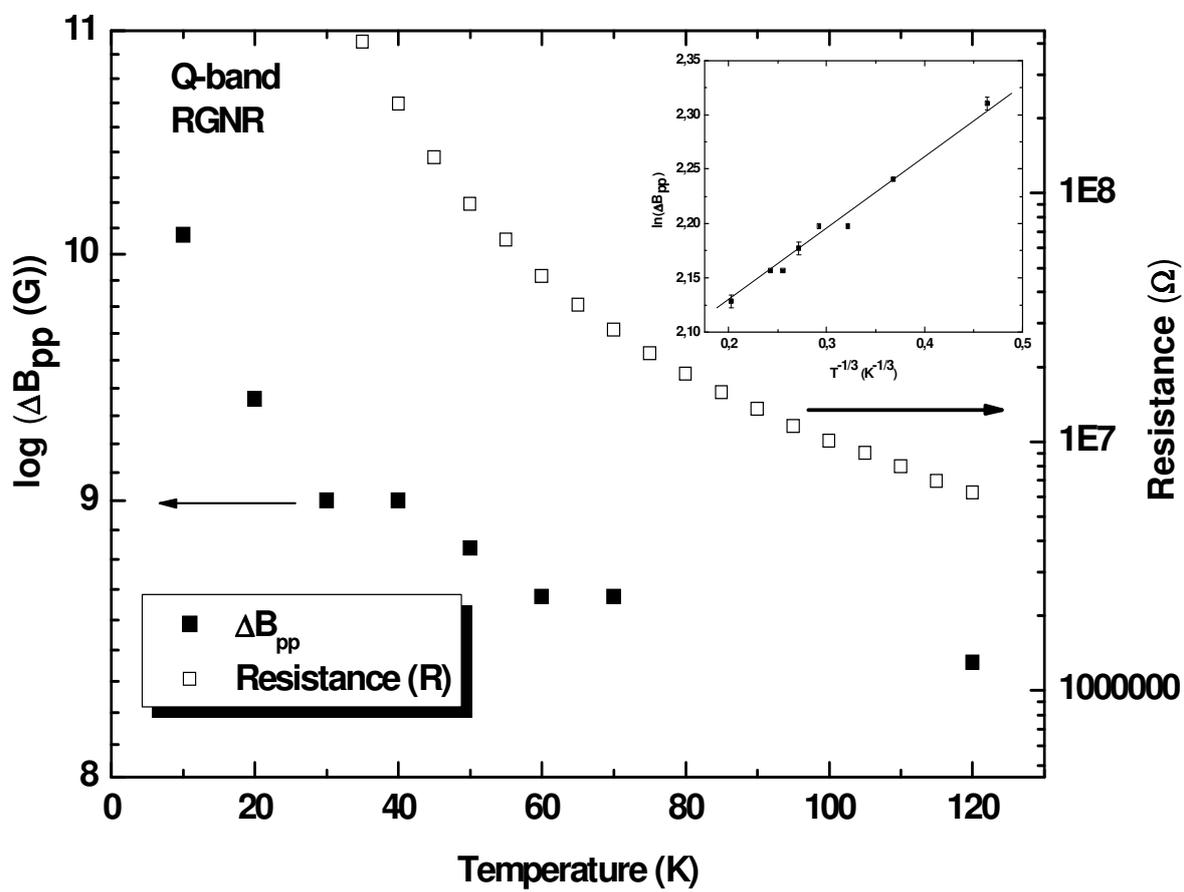

**Fig. 4**